\documentclass[a4paper]{jpconf}

\usepackage{multirow}
\usepackage{dcolumn}
\usepackage{amsmath,amssymb}
\usepackage{amsfonts}
\usepackage{stmaryrd}
\usepackage{mathrsfs}

\usepackage{fontenc}
\usepackage{times}
\usepackage{mathptmx}

\usepackage{graphicx}
\usepackage{bm}
\usepackage[english]{babel}

\begin{document}

\newcommand{\Refs}{Refs.~}
\newcommand{\Ref}{Ref.~}
\newcommand{\Sec}{Section~}
\newcommand{\Tab}{Table~}
\newcommand{\Eq}{Eq.~}
\newcommand{\bld}[1]{\boldsymbol{#1}}

\def\nuebar{{\rm \bar{\nu}_e}}
\def\nuebare{{\rm \bar{\nu}_{e}-e}}
\def\nue{{\rm \nu_e}}
\def\nuee{{\rm \nu_{e}-e}}
\def\s2tw{{\rm sin ^2 \theta_{W}}}
\def\sqs2tw{\rm{sin ^4 \theta_{W}}}
\def\munu{{\rm \mu_{\nu}}}
\def\munubar{\rm{\mu_{\bar{\nu}_{e}}}}
\def\cpkkd{\rm{kg^{-1} keV^{-1} day^{-1}}}
\def\nuchrad{\rm{\langle r_{\bar{\nu}_e}^2\rangle}}
\def\nocrv{\rm{\hspace*{0.2cm} \nearrow \hspace*{-0.6cm} CRV }}

\def\er{{\rm \varepsilon_{ee}^{eR}}}
\def\el{{\rm \varepsilon_{ee}^{eL}}}
\def\elr{{\rm \varepsilon_{ee}^{eL,R}}}
\def\etl{{\rm \varepsilon_{e\tau}^{eL}}}
\def\etr{{\rm \varepsilon_{e\tau}^{eR}}}
\def\etlr{{\rm \varepsilon_{e\tau}^{eL,R}}}
\def\eml{{\rm \varepsilon_{e\mu}^{eL}}}
\def\emr{{\rm \varepsilon_{e\mu}^{eR}}}
\def\emlr{{\rm \varepsilon_{e\mu}^{eL,R}}}

\def\Lambdaup{\Lambda_{U}}
\def\dsca{d_{S}}
\def\dvec{d_{V}}

\title{Final results of $\nuebare$ scattering cross-section
measurements and constraints on new physics}

\author{Muhammed Deniz $^{a,b,c}$,
Selcuk Bilmis $^{b,c}$ and Henry T. Wong $^{c}$\\
(on behalf of the TEXONO Collaboration)}

\address{$^a$ Department of Physics, Karadeniz Technical University,
        Trabzon 61080, Turkey.\\
$^b$ Department of Physics, Middle East Technical University,
Ankara 06531, Turkey.\\
        $^c$ Institute of Physics, Academia Sinica, Taipei 11529, Taiwan.\\}
\ead{htwong@phys.sinica.edu.tw}

\begin{abstract}

The $\nuebare$ elastic scattering cross-section was
measured with a CsI(Tl) scintillating crystal detector array with a
total mass of 187~kg at the Kuo-Sheng Nuclear Power Station.
The detectors were exposed to a reactor $\bar{\nu}_{e}$ flux of
$\rm{6.4\times 10^{12} ~ cm^{-2}s^{-1}}$ originated from a core with
2.9~GW thermal power.
Using 29882/7369 kg-days of Reactor ON/OFF
data, the Standard Model (SM) of electroweak interaction was probed
at the 4-momentum transfer range of
$\rm{Q^2 \sim 3 \times 10^{-6} ~ GeV^2}$.
A cross-section ratio of $\rm{R_{expt}=[1.08 \pm
0.21~(stat) \pm 0.16~(sys)] \times R_{SM} }$ was measured.
Constraints on the electroweak parameters $\rm{(g_V,g_A )}$ were
placed, corresponding to a weak mixing angle measurement of
$\rm{ \s2tw=0.251 \pm 0.031~(stat) \pm 0.024~(sys)} $.
Destructive interference in the SM $\nuebar$-e processes
was verified.
Bounds on neutrino anomalous
electromagnetic properties
(neutrino magnetic moment and neutrino charge radius),
non-standard neutrino interactions, 
upparticle physics and
non-commutative physics
were placed.
We summarize the experimental details and results,
and discuss projected sensitivities with realistic
and feasible hardware upgrades.

\end{abstract}

Neutrino-electron scattering is pure leptonic process which is
exactly evaluated with electroweak theory.
Cross-section measurement of
$\nuebare$ scattering with reactor neutrinos~\cite{texononue}
provides a new window to study
the Standard Model (SM)
electroweak parameters as well as
anomalous neutrino electromagnetic properties
such as charge radius and magnetic moments
in a different kinematics regime from that at accelerator.
In addition,
new physics beyond SM such
as Non-Standard neutrinos interaction (NSI) and Unparticle
Physics (UP)~\cite{texononsi},
as well as the scale ($\Lambda_{NC}$) for
Non-commutative Physics~\cite{texononcom}
can be probed.
The $\nuebare$ scattering cross-section
in the laboratory frame is:
\begin{equation}
\left[ \frac{d\sigma}{dT}(\bar{\nu}_{e}e ) \right] _{SM}  =
\frac{G_{F}^{2}m_{e}}{2\pi }  \cdot
[ ~ \left(g_{V}-g_{A}\right) ^{2}
 +  \left( g_{V}+g_{A}+2\right) ^{2}\left(1-
\frac{T}{E_{\nu }}\right) ^{2}
 -  (g_{V}-g_{A})(g_{V}+g_{A}
+2)\frac{m_{e}T} {E_{\nu}^{2}}  ~ ] ,
\label{eq::gvga}
\end{equation}
where $G_{F}$ is the Fermi constant,
T is the kinetic energy of the recoil electron, $E_\nu$ is the
incident neutrino energy, $m_e$ is mass of the electron
and $g_{V}$, $g_{A}$ are, respectively, the
vector and axial-vector coupling constants.
The SM assignments to the coupling constants
are:
$ g_{V}=-\frac{1}{2}+2\sin ^{2}\theta _{W}$ and
$g_{A}=-\frac{1}{2}$,
where $\s2tw$ is the weak mixing angle.
Interactions of $\nuebare$ and $\nu - e$ are two of the few
SM processes which proceed via charged- and neutral-currents
{\it and} their interference, such that the event rates ($R$)
can be expressed by its components as:
$R_{expt} = R_{CC} + R_{NC} + \eta \cdot R_{Int}$.


A model-independent formulation of NSI
is to characterize $\nu - e$ scatterings as
four-Fermi interactions with new
couplings  $\epsilon_{\alpha\beta}^{eP}$ which
describe the coupling strength
with respect to $G_F$.
The helicity states are denoted by P (=L,R),
and  $(\alpha , \beta)$ stand for
the lepton flavor (e, $\mu$ or $\tau$)
for the incoming/outgoing neutrinos.
The cases where
$\alpha = \beta$ and $\alpha \ne \beta$
correspond to Non-Universal (NU)
and Flavor-Changing (FC) NSI, respectively.
The $\nuebare$ cross-section including both SM and NSI
is given by~\cite{nsi}:
\begin{eqnarray}
\left[\frac{d\sigma}{dT}\right]_{SM+NSI} = ~~\frac{2 G_F^2 m_e}{\pi}
\cdot [~\left(\tilde g_R^2 + \sum_{\alpha \neq e}|\epsilon_{\alpha
e}^{e R}|^2 \right) + \left((\tilde g_L + 1)^2 + \sum_{\alpha \neq
e}|\epsilon_{\alpha e}^{e L}|^2 \right)
\left(1 - \frac{T}{E_{\nu}}\right)^2 \nonumber \\
- \left(\tilde g_R(\tilde g_L + 1)+\sum_{\alpha \neq
e}|\epsilon_{\alpha e}^{e R}||\epsilon_{\alpha e}^{e L}|
\right)\frac{m_e T}{E^2_{\nu}}] ~~~,
\label{eq_nsics}
\end{eqnarray}
where $\tilde g_L=g_L+\el$ and $\tilde g_R=g_R+\er$, defined in
terms of the chiral couplings:
$g_L = \frac{1}{2}(g_V+g_A) = -\frac{1}{2}+\s2tw$
and
$ g_R = \frac{1}{2}(g_V-g_A) = \s2tw \label{eq::glr}$.

The measurement of $\nuebare$ elastic scattering at
$Q^2 \sim {\rm MeV^2}$
was performed by the TEXONO Collaboration
at the Kuo-Sheng Reactor Neutrino Laboratory in Taiwan.
The experimental details were documented
in Refs.~\cite{texononue,texonocsi}.
The CsI(Tl) target crystals were configured
as a $12\times 9$ array inside a shielding structure
with $\sim$50~tons of materials enclosed
by an active cosmic-ray scintillator veto (CRV) system.
Each single crystal module has a
hexagonal-shaped cross-section with 2~cm side, 40~cm length and
modular mass of 1.87~kg.
The detector consisted of 100 crystals giving
a total mass of 187~kg.
The light output was read out at both ends
of the crystal by photomultipliers (PMTs)
with low-activity glass of 29~mm diameter.
The PMT signals were recorded by 20~MHz Flash Analog-to-Digital-Converters
(FADCs) running on a VME-based data acquisition system. The sum of the two
PMT signals gives the energy of the event, while their difference provides
information on the longitudinal ``Z'' position. An energy resolution of
$4\%$ and a Z-resolution of $<$1 cm RMS at 660~keV as
well as excellent $\alpha$/$\gamma$ event identification by pulse shape
discrimination (PSD) were demonstrated.

\begin{figure}[hbt]
\begin{minipage}{18pc}
\includegraphics[width=18pc]{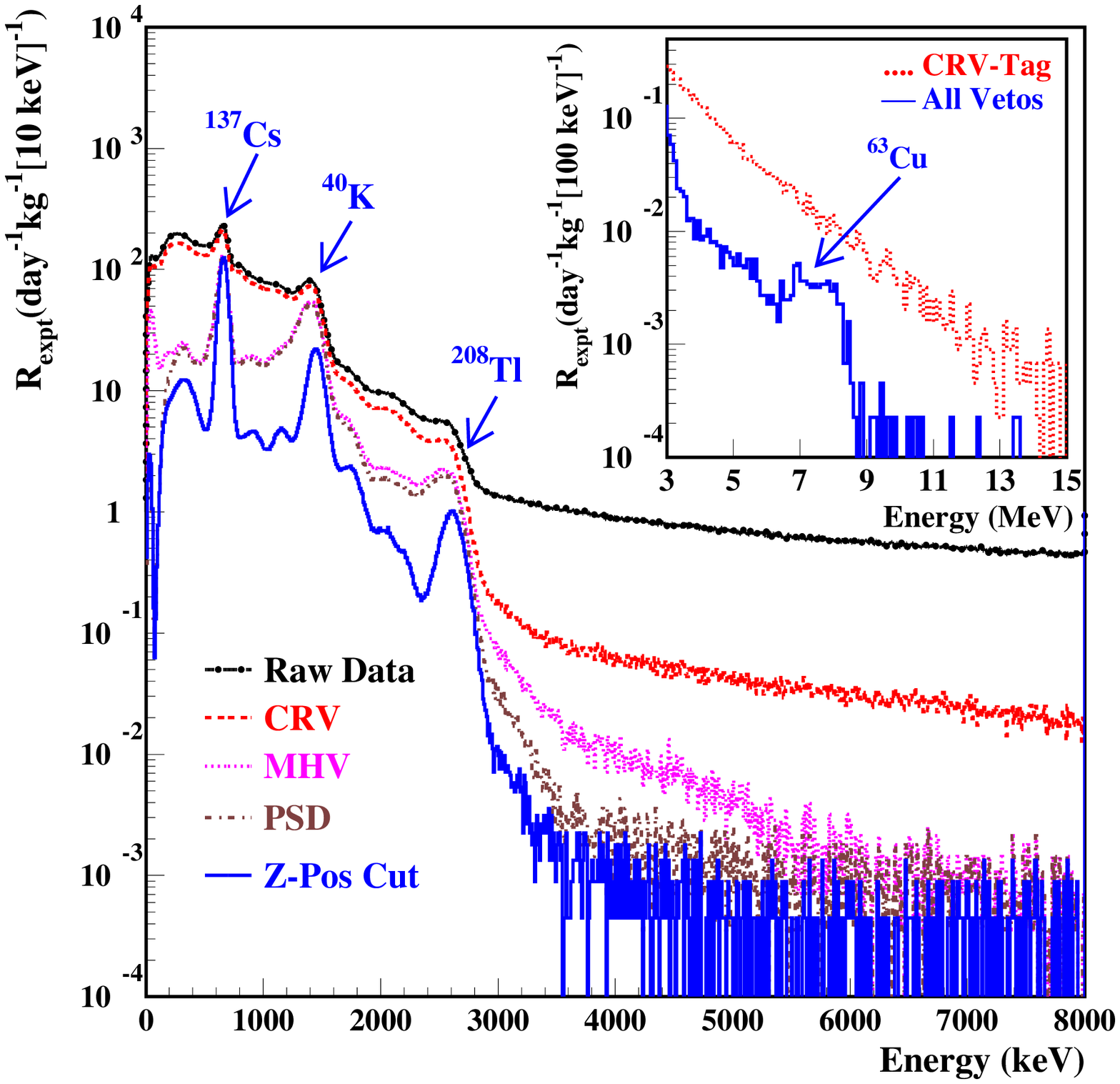}
\caption{\label{fig_cutspec}
The measured energy spectra
at various stages of the analysis
showing the effects of successive selection
cuts.
}
\end{minipage}\hspace{2pc}%
\begin{minipage}{18pc}
\includegraphics[width=18pc]{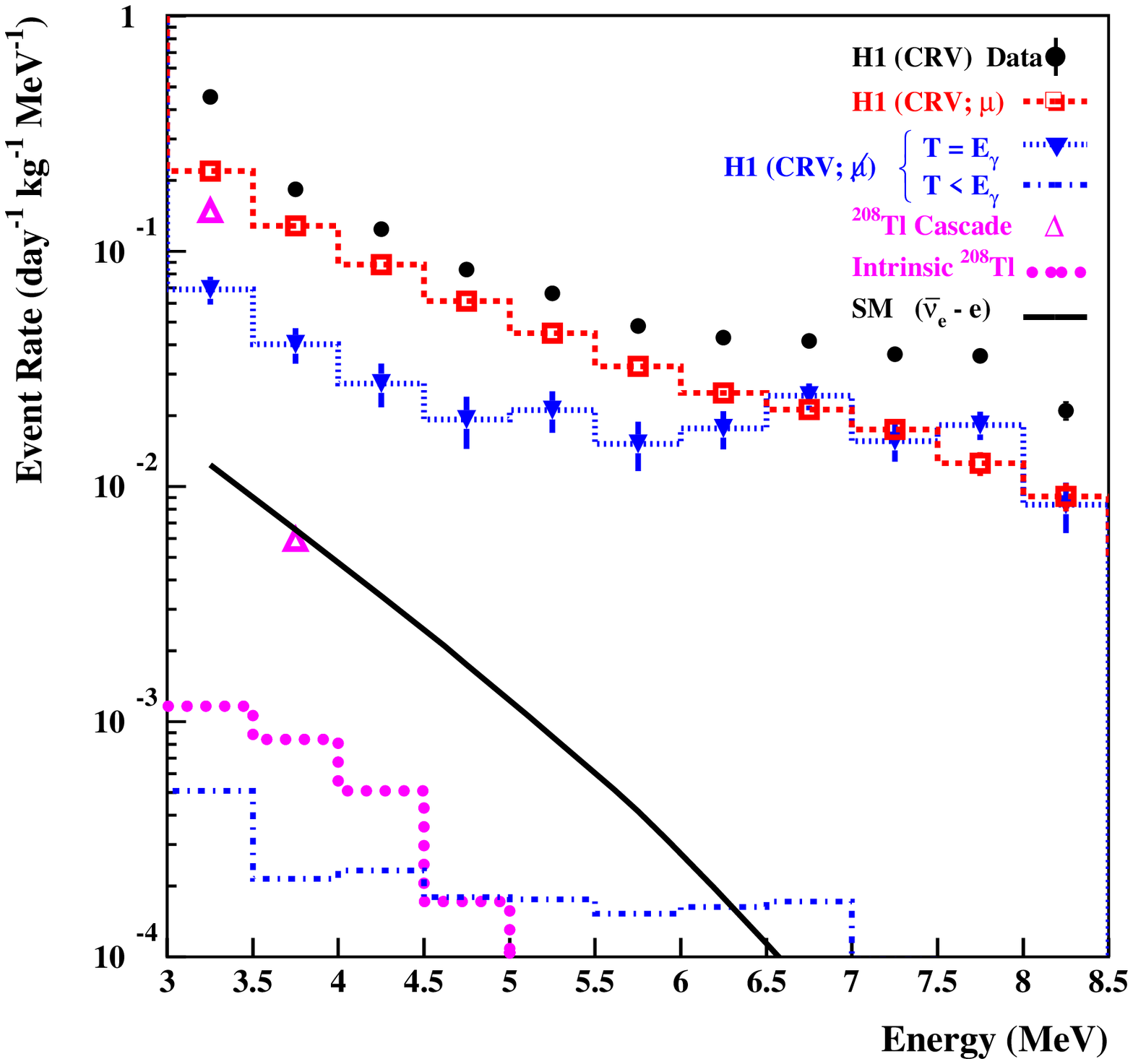}
\caption{\label{bkgchannel}
Measured H1 spectrum and
the different evaluated background channels.
The SM $\nuebar -$e contributions are
overlaid.
}
\end{minipage}
\end{figure}

Neutrino-induced candidate ``single-hit (H1)'' events
were selected through the
event-by-event suppression in the various background channels:
(a) cosmic-ray induced events by CRV,
(b) $\gamma$-induced anti-Compton events by
multiplicity veto (MHV),
(c) accidental and $\alpha$- events by PSD,
and (c) external background by longitudinal Z-position cut.
Residual H1 background channels at the relevant 3$-$8~MeV range
are depicted in Figure~\ref{bkgchannel},
where the dominant components include: (i) cosmic-ray induced
events where CRV are missing; (ii) ambient high energy
$\gamma$-rays following neutron capture of $^{63}$Cu;
and (iii) cascade $\gamma$-rays from $^{208}$Tl.
Their contributions were evaluated from
{\it in situ} measurement of CRV inefficiencies,
multi-hit samples and the $^{208}$Tl-2614~keV lines,
coupled with simulation studies.
and the results provide the
second background measurement.
The combined background (BKG) were statistically
subtracted from the H1 spectra.

\begin{figure}[hbt]
\begin{minipage}{18pc}
\includegraphics[width=18pc]{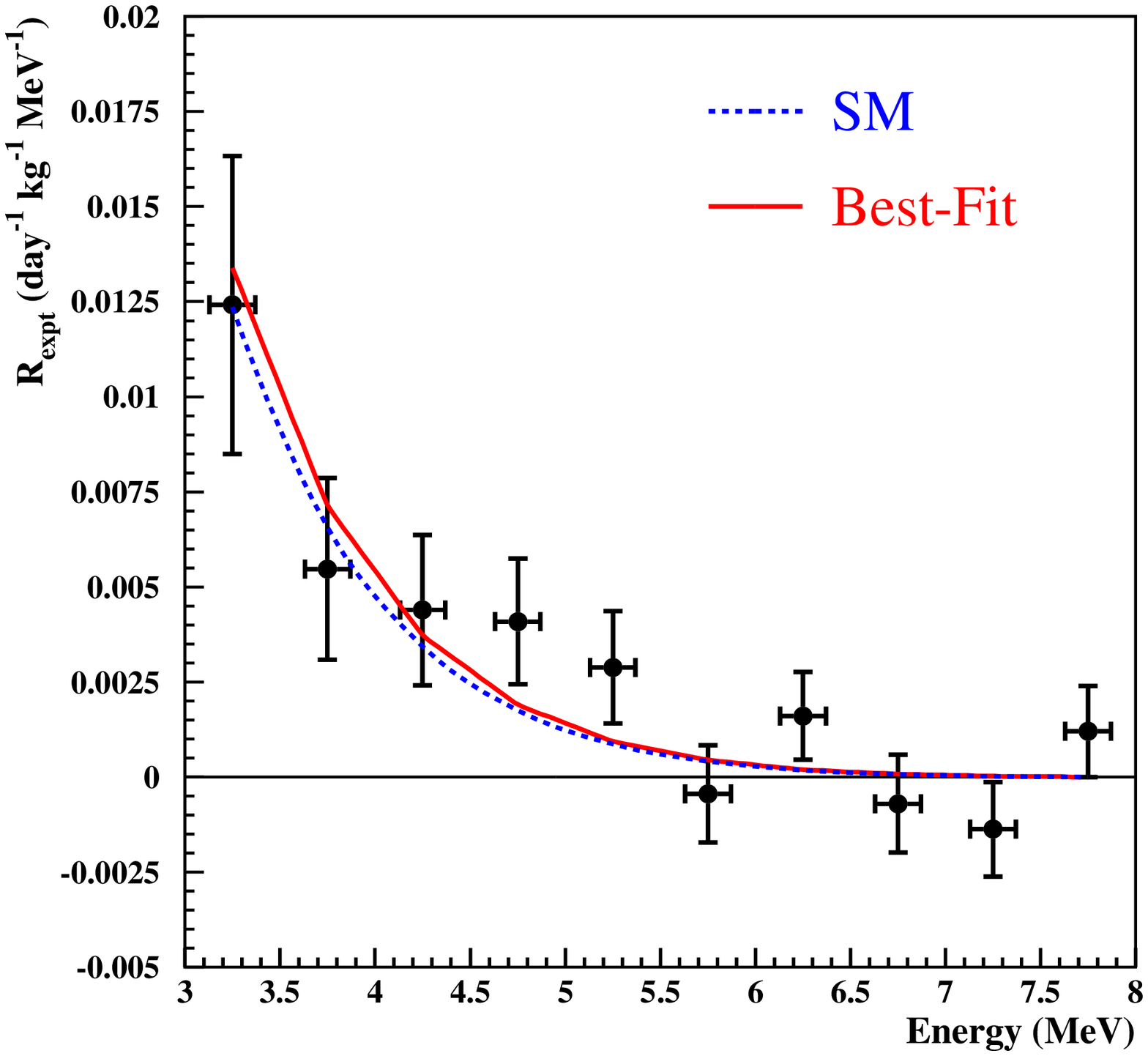}
\caption{\label{residual} The combined ON-BKG residual spectrum
together with those of SM and best-fit results.}
\end{minipage}\hspace{2pc}%
\begin{minipage}{18pc}
\includegraphics[width=18pc]{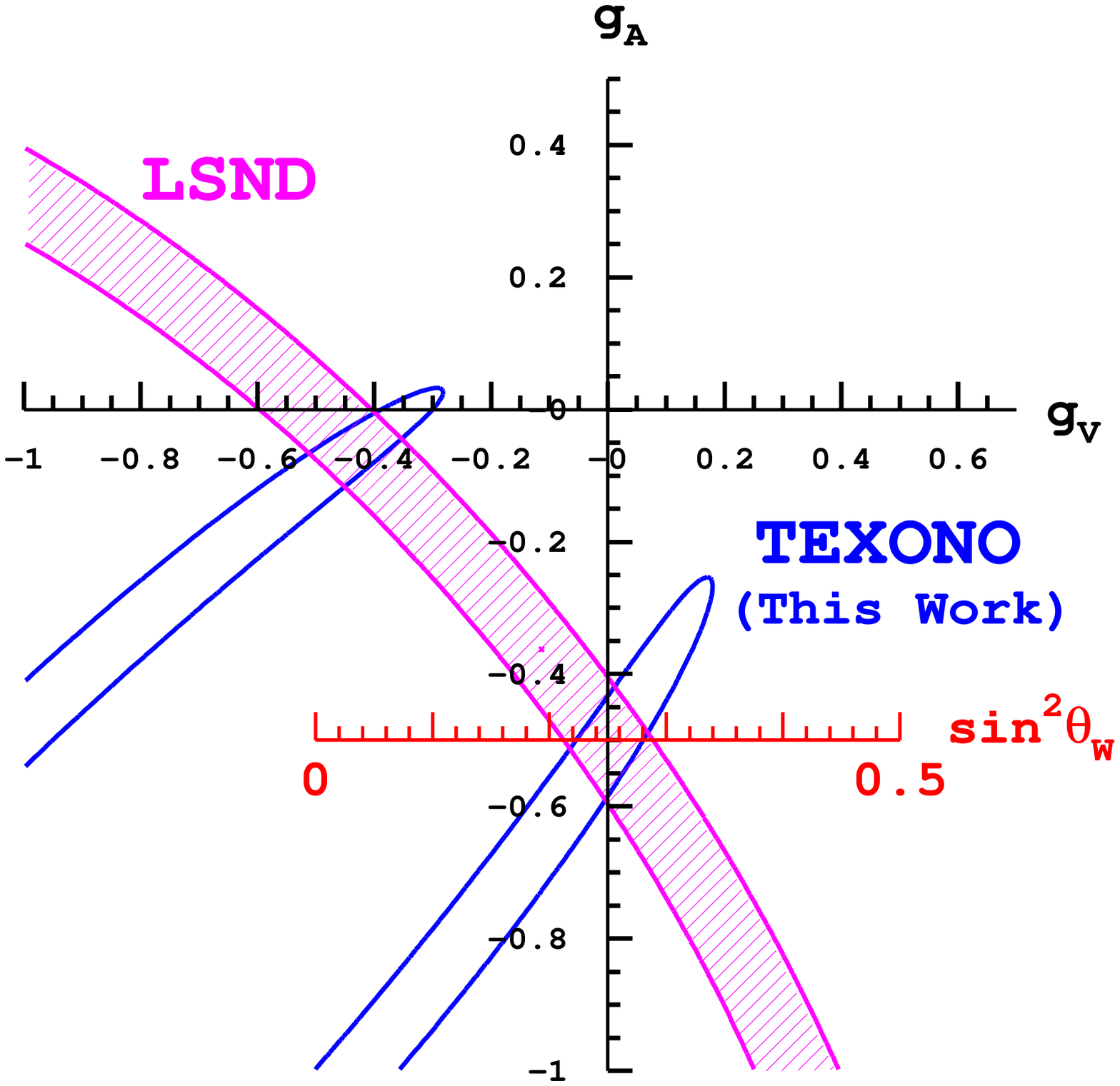}
\caption{\label{gvga} The 1$\sigma$ allowed region in $ g_{V}-g_{A}$
space, and in the $\s2tw$ axis, for both $\nuebare$ (TEXONO) and
$\nue - e$ (LSND) scattering experiments.

}
\end{minipage}
\end{figure}

A total of 29882/7369~kg-day of Reactor ON/OFF data was recorded.
The $OFF-BKG$ spectrum is consistent with zero showing the
background modeling and subtraction is valid.
The combined $ON-BKG$ residual spectrum is displayed in
Figure~\ref{residual}, from which various electroweak parameters
were derived.
The excess in the residual spectrum corresponds to
$\sim$414 neutrino-induced events,
and an event-rate relative to that of SM at
a ratio of
$\xi \equiv R_{expt}/R_{SM} =
[1.08 \pm 0.21 (stat) \pm 0.16 (sys)]$ which implies
$\s2tw = 0.251 \pm 0.031 (stat) \pm 0.024 (sys)$.
The allowed region in the $g_V - g_A$
plane is depicted in Figure~\ref{gvga}. The accuracy improves over
previous reactor $\nuebare$ experiments~\cite{texononue} as well as that in
accelerator-based $\nue - e$ scattering
experiments~\cite{lampf}.
The measured interference parameter of  $\eta = -0.92 \pm 0.30 (stat) \pm 0.24 (sys)$
verifies the SM prediction of destructive interference ($\eta_{SM}= -1$).

Existence of neutrino magnetic moment ($\munubar$) would contribute
an additional term to the cross-section of Eq.~\ref{eq::gvga}:
\begin{equation}
\left( \frac{d\sigma }{dT}\right) _{\mu _{\nu}}=\frac{\pi \alpha
_{em}^{2}\mu _{\nu}^{2}}{m_{e}^{2}}\left[
\frac{1-T_{e}/E_{\nu}}{T_{e}}\right] \label{eq_mm} ~~,
\end{equation}
where $\alpha_{em}$ is the fine structure constant.
From a best-fit analysis, a limit of
$\munubar < 2.2 \times 10^{-10} \times \mu_{B}$
at 90\% confidence level (CL) was derived. A finite neutrino charge radius
$\nuchrad$ would lead to radiative corrections which modify the
electroweak parameters by
$g_{V}  \rightarrow -\frac{1}{2}+2\s2tw + (2\sqrt{2}\pi\alpha_{em}/3G_F) \nuchrad$
and $\s2tw \rightarrow \s2tw + (\sqrt{2}\pi\alpha_{em}/3 G_F)\nuchrad$.
The allowed range at 90\% CL is
$-2.1 \times 10^{-32} < \nuchrad < 3.3 \times 10^{-32}$.
New constraints on NSI and UP couplings~\cite{texononsi}
as well as on $\Lambda_{NC}$~\cite{texononcom} 
were also derived,
as displayed in Table~\ref{tab_eelr}
for those of NSI.

\begin{table} [hbt]
\caption{\label{tab_eelr} Constraints at 90\% CL
on the NSI couplings. The projected sensitivities
correspond to improved features listed in Table~\ref{tab_prosensit}.
}
\begin{tabular}{cccccc} \hline
\multicolumn{2}{c}{NSI} & Measurement & & Bounds & Projected \\
\multicolumn{2}{c}{Parameters} & Best-Fit$\pm$stat.$\pm$sys &
$\chi ^2$/dof & at 90\%
C.L. & Sensitivities  \\ \hline
\multirow{2}*{NU \{} & $\el$
& $\el =0.03 \pm 0.26 \pm 0.17$ & 8.9/9 &  $-1.53 < \el < 0.38 $  &
$\pm$0.015 \\

& $\er$
&  $\er =0.02 \pm 0.04 \pm 0.02$ &  8.7/9  & $-0.07 < \er < 0.08 $  &
$\pm$0.002 \\ \hline

\multirow{2}*{FC \{} & $\eml,\etl$ & $\eml ^2 ( \etl ^2 ) = 0.05 \pm
0.27 \pm 0.24$ & 8.9/9  & $|\eml|(|\etl|) < 0.84$   &
$\pm$0.052 \\

& $\emr,\etr$ & $\emr ^2 ( \etr ^2 ) = 0.008 \pm 0.015 \pm 0.012$ &
8.7/9  & $|\emr|(|\etr|) < 0.19$   & $\pm$0.007 \\ \hline
\end{tabular}
\end{table}

\begin{table}[hbt]
\caption{
Projected statistical sensitivities
on $\xi$ and $\s2tw$
under various realistically
achievable improvement to
the experiment.
}
\label{tab_prosensit}
\begin{center}
\begin{tabular}{lcc} \hline
Improvement& $\Delta_{stat} ( \xi )$  & $\Delta_{stat} [ \s2tw ]$  \\ \hline
This Work &  0.21 & 0.031 \\
\multicolumn{3}{l}{\underline{Improved Feature} :} \\
~ A. $\times 30$ Data Strength & 0.038 & 0.0057 \\
~ B. Background Reduction & & \\
~~~ B1: $>$99\% Cosmic-Ray Efficiency & 0.12 & 0.018 \\
~~~ B2: $\times \frac{1}{10}$ Reduction in 
Ambient \& $^{208}$Tl $\gamma$'s & 0.16 & 0.024 \\
~ $\ast$ With Both B1+B2 & 0.05 & 0.007 \\
All Features A+B1+B2 Combined  & 0.009 & 0.0013 \\ \hline
\end{tabular}
\end{center}
\end{table}

The sensitivities can be further enhanced in
future experiments.
As illustrations,
the projected improvement under various realistically
achievable assumptions are summarized in
Table~\ref{tab_prosensit}.
Electromagnetic calorimeters using CsI(Tl) with
tens of tons of mass have been constructed, such that
the target mass is easily expandable.
As shown in Figure~\ref{bkgchannel},
the dominant background above 3~MeV were all external
to the target scintillator.
Accordingly, they will be attenuated
effectively through self-shielding in a target with
bigger mass.
The incorporated features
listed in Table~\ref{tab_prosensit}
correspond to 30 times increase
in data strength (for instance, with 1~ton fiducial mass
and 1000~days data taking)
and $\times$1/10 suppression in background.
The statistical accuracies
can be improved to
0.8\% and 0.12\%
for $\xi$ and $\s2tw$, respectively.
Systematic uncertainties originate mainly from
the evaluation of the reactor neutrino spectra.
This can be overcome by a simultaneous measurement
of the $\nuebar$ spectra via the matured
inverse beta-decay process
$\nuebar + p \rightarrow n + e^+$ with, for instance,
large liquid scintillator detectors.

\end{document}